\documentclass[aps,twocolumn, superscriptaddress, english, nofootinbib, 10pt]{revtex4-2}

\usepackage{float}

\usepackage{amssymb, amsmath}
\usepackage[utf8]{inputenc}
\usepackage{subfigure}
\usepackage{comment}
\usepackage{graphicx}
\usepackage{hyperref}
\usepackage{enumerate}
\usepackage{lipsum}
\usepackage{xspace}
\usepackage{aas_macros}
\usepackage{enumitem}
\usepackage{booktabs}

\usepackage[usenames,dvipsnames]{xcolor}
\hypersetup{
    colorlinks = true,
    citecolor = {MidnightBlue},
    linkcolor = {BrickRed},
    urlcolor = {BrickRed}
}

\newcommand{\nv}{\hat{\bf n}}
\newcommand{\lv}{\hat{\bf e}_\theta}
\newcommand{\mv}{\hat{\bf e}_\varphi}

\newcommand{\rot}[1]{{\sf R}_{#1}}
\newcommand{\planck}{{\sl Planck}\xspace}

\begin{document}

\title{Dipoles for everyone: the pseudo-$C_\ell$ approach to directional stacking}

\author{Lea Harscouet}
\email{lea.harscouet@physics.ox.ac.uk}
\affiliation{Department of Physics, University of Oxford, Denys Wilkinson Building, Keble Road, Oxford OX1 3RH, United Kingdom}
\author{Amy Wayland}
\affiliation{Department of Physics, University of Oxford, Denys Wilkinson Building, Keble Road, Oxford OX1 3RH, United Kingdom}
\author{David Alonso}
\affiliation{Department of Physics, University of Oxford, Denys Wilkinson Building, Keble Road, Oxford OX1 3RH, United Kingdom}

\begin{abstract}
  Stacking (i.e. averaging) the value of a given astrophysical field around sources allows us to detect new cosmological signatures, such as the kinematic Sunyaev-Zel'dovich, and gain insight on the astrophysical properties of galaxies and their environment. Further information may be gained by orienting these stacks along preferred axes defined by a local directed field, such as the transverse galaxy velocities, galaxy shapes, or the local tidal forces. Examples of this are searches for the moving lens effect, the detection of dipole signatures, or the study of cosmic filaments. Here we show that all directional stacking signals may be reconstructed, without loss of information, in terms of the cross-power spectrum between the quantity of interest and the $E$ and $B$ modes of the spin field used to define the preferred axes weighted by the local galaxy density. The power spectrum approach has several practical advantages, in terms of speed, finite-resolution effects, data visualisation, and combination with other cosmological probes. We also argue that, in some cases, such as stacking using velocities or tidal forces reconstructed from the density field, the recovered signal may be dominated by information that is already present in the cross-spectrum between the target field and the galaxy overdensity itself.
\end{abstract}

\maketitle

\section{Introduction}\label{sec:intro}
  The stacking of local patches of cosmological fields around sources is commonly used to extract signals which would otherwise be too faint to detect for each individual object. This technique can significantly boost the signal-to-noise ratio (SNR) by enhancing a shared signal across $N$ patches by $N$, while mitigating uncorrelated sources of noise -- which only scale as $\sqrt N$. As a result, stacking enables detections of coherent signals about $\sqrt N$ times more significant than any individual measurement.
  A typical example of this stacking measurement is that of the kinematic Sunyaev-Zeldovich (kSZ) effect, in which CMB temperature maps are stacked at the positions of galaxies, weighted by an estimate of the local radial velocity field -- allowing us to probe baryonic physics in the outskirts of galaxies \citep{liMatchedFilterOptimization2014,1510.06442,2009.05557,2305.06792,2407.07152,2503.19870}.

  One can also look at higher-order multipole patterns in the fields of interest around sources: this is done by first aligning the patches with a preferred direction along which one would expect to find an enhanced signal. This so-called \textit{oriented} or \textit{directional} stacking is often used to measure cosmological signatures along directions given by e.g. moving sources or density ridges. For instance, \cite{caiDetectionCosmologicalDipoles2025} detected, through directional stacking, various dipolar signatures aligned with the reconstructed transverse velocities of galaxies, imprinted on the CMB lensing and temperature fields and the galaxy density field itself. The Integrated Sachs-Wolfe (ISW) effect, as well as its small-scale manifestation, the Moving Lens (ML) effect \citep{hotinliTransverseVelocitiesMatter2023,beheshtiMovingLensEffect2025,hotinliMLinprep}, could thus be detected at high significance in the CMB temperature field by first aligning each temperature subsets along the transverse velocity of their respective source galaxy. These dipole measurements could then be used to constrain the cosmological parameters governing structure growth.
  
  Directional stacking can also prove useful to detect the presence of low-density gas along the filamentary structures of the cosmic web (e.g. via emission lines \cite{gallegoStackingCosmicWeb2018}) or even potential magnetic fields (via synchrotron radiation and X-ray emission \cite{vernstromDiscoveryMagneticFields2021,hodgsonStackingSynchrotronCosmic2022}). The kinematic and thermal Sunyaev-Zeldovich signatures of filaments was also observed in e.g. \cite{hadzhiyskaTracingCosmicGas2024, tanimuraSearchWarmHot2019, degraaffProbingMissingBaryons2019, lokkenSuperclusteringAtacamaCosmology2022}, and the rotational kinematic SZ (rkSZ) effect was detected in \cite{baxterConstrainingRotationalKinematic2019, goldsteinEvidenceGalaxyCluster2025, yangMeasurementGasRotation2026} by stacking CMB temperature maps oriented along the rotational spin of galaxy clusters. Recently, \cite{nikjooWeakLensingPhotometric2026} have also detected a shear signal produced by foreground filamentary structures using oriented stacking. Although most applications of this technique are relevant for the study of large-scale structures, one of the early applications of directional stacking was in fact developed for CMB studies \cite{planckcollaborationPlanck2015Results2016} in order to check for small-scale anisotropies in CMB temperature and polarization fields around hot spots. 

  Real-space stacking -- oriented or otherwise -- provides an intuitive visual representation of the signal, as it is essentially the real-space equivalent of the correlation function of the fields involved \cite{caiDetectionCosmologicalDipoles2025, planckcollaborationPlanck2015Results2016}, and can be interpreted as the average profile of a given astrophysical field around cosmological structures. Such parallels can be drawn between traditional plots of the baryon acoustic oscillation (BAO) ring, revealed by stacking the reconstructed galaxy field at the position of sources \cite{padmanabhan2DistanceZ0352012}, and its peak in the corresponding correlation function -- or its Fourier space equivalent, the BAO wiggles seen in the galaxy power spectrum. Mathematically, any real-space stacking measurement should be mappable to a power spectrum measurement. For instance, this stacking-power spectrum connection was recently formalised for the traditional kSZ stacking estimator \cite{harscouetKSZEveryonePseudoCl2025, quPrecisionKinematicSunyaevZeldovich2026, hadzhiyskaPrecisionKinematicSunyaevZeldovich2026}, in terms of the cross-power spectrum between the CMB temperature map and the radial galaxy momentum field. The power spectrum alternative to stacking offers several advantages: power spectrum uncertainties are largely uncorrelated, facilitating ``$\chi^2$-by-eye'' tests, and can be estimated accurately and analytically. Power spectra are often more closely related to first-principles cosmological predictions, and their use also allows for the adoption of common analysis choices in multi-probe studies (e.g. consistent model choices and scale cuts). Finally, power spectrum estimators are typically faster than real-space correlations, and the advent of catalogue-based estimators \cite{2312.12285,2407.21013}, based on irregular-grid spherical harmonic transforms \cite{2304.10431}, allows for measurements on arbitrarily small scales avoiding finite-resolution effects.

  If regular stacking finds a mathematical equivalent in a power spectrum measurement, the directional stacking approach similarly must have a harmonic-domain counterpart. This paper focuses on providing a framework for expressing most directional stacking estimators as a power spectrum measurement. Specifically, we show that the information encoded in directional stacking is equivalent to taking the $E$-mode power spectrum of a spin-$s$ field (which carries information about the preferred direction) and the background field (which carries the physical imprint). Interestingly, the spin of the field is indicated by the type of anisotropy observed in real-space stacking -- $s=1$ for a dipolar signature, $s=2$ for a quadrupolar signature, and so on. A similar framework is also proposed in the work of \cite{hotinliMLinprep}.
  
  In Section \ref{sec:math}, we lay out the formalism and provide an expression for directional stacking in which all the cosmologically relevant information comes from a power spectrum. In Section \ref{sec:ex}, we demonstrate this equivalence numerically using mock data. We also showcase the application of the harmonic-space formalism to real data. First, we reproduce the dipole measurements from \cite{caiDetectionCosmologicalDipoles2025}, where the spin field of interest is the galaxy transverse velocity field ($s=1$). Secondly, we measure the quadrupole signature of the thermal SZ effect around filaments as investigated in e.g. \cite{lokkenSuperclusteringAtacamaCosmology2022}, where the spin field is informed by the reconstructed tidal tensor ($s=2$). We also discuss the usefulness of directional stacking when the directing field is reconstructed from the galaxy overdensity itself. We summarise our main results and conclude in Section \ref{sec:conc}.
  
\section{Directional stacking and power spectra}\label{sec:math}
  In this section we derive the connection between directional stacking and angular power spectra of spin fields. Some of the mathematical formalism used here to define and treat rotations on the sphere can be found in Appendix \ref{app:WignerDmatrices}.

  \subsection{Directional stacking}\label{ssec:math.stack}
\begin{figure}
      \centering
      \includegraphics[width=0.95\linewidth]{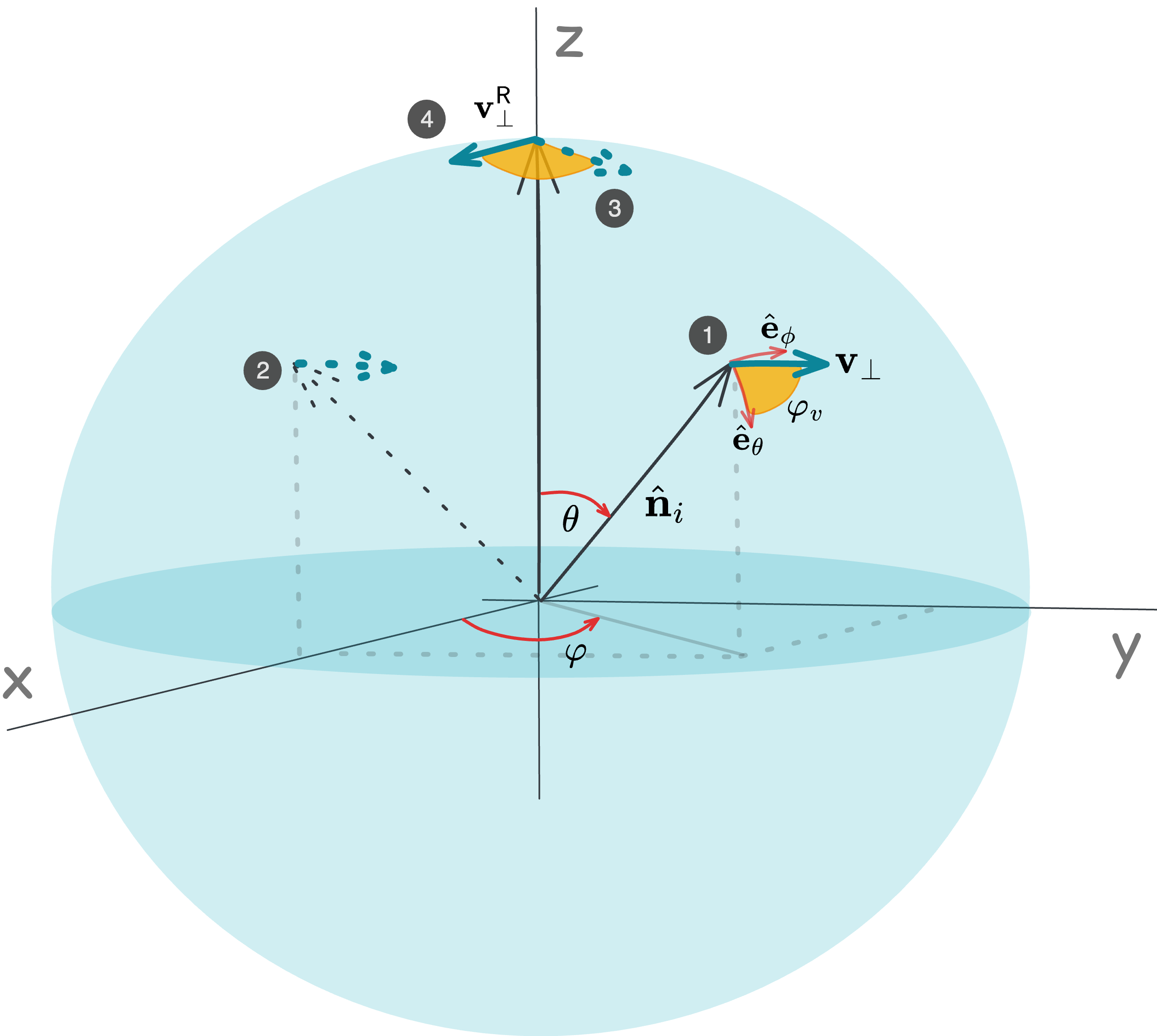} 
      \caption{Rotation steps. The original galaxy is placed at $\nv_i = (\theta, \varphi)$ (1), and is moved to (2) after a $-\varphi$ rotation about the $z$ axis. In (3) is the position of the galaxy after a second rotation of $-\theta$ about the $y$ axis, and in (4) its position after a further $-\varphi_v$ rotation about the $z$ axis. These successive rotations can be written as $D(-\varphi_v, -\theta, -\varphi)$ in the $z-y-z$ convention.}
      \label{fig:spheres_setup}
    \end{figure}

    Let $X(\nv)$ be an external continuous field defined on the sphere that we wish to stack (e.g. a CMB temperature or lensing convergence map), and let $\nv_i$ be the angular coordinates of a set of sources on which we will stack. Furthermore, let ${\bf v}_i\equiv(v_{1,i}, v_{2,i})$ be a spin-$s$ field measured at the positions of each source, with two components $v_1$ and $v_2$. This field determines a preferred direction, parametrised by the angle $\varphi_v$, defined in terms of the complex phase of the spin field: $\tilde{v}\equiv v_1+iv_2=v\,e^{is\varphi_v}$, where $v\equiv\sqrt{v_1^2+v_2^2}$. $\varphi_v$ is the angle this preferred direction makes with respect to the locally-defined $x$-axis. Here we adopt the usual CMB convention, in which the local $x$ and $y$ axes are aligned with the directions of increasing colatitude $\theta$, and azimuth $\varphi$, respectively (i.e. with the unit vectors $\lv$ and $\mv$ -- see Fig. \ref{fig:spheres_setup}).
    
    To obtain a more intuitive picture of this set-up, let us consider the most-common example of directional stacking, in which ${\bf v}\equiv{\bf v}_\perp$ is the transverse peculiar velocity of the source. In this case $v_1$ and $v_2$ are the components of this vector along $\lv$ and $\mv$, and $\varphi_v\equiv{\bf v}\cdot\lv/v$.

    With this setup, the stacking estimator can be described by the following steps:
    \begin{enumerate}
      \item \textbf{Rotation step}. For each galaxy position $\nv_i$, we rotate the $X$ map such that $\nv_i$ is at the North pole, and that $\tilde{v}_i$ is aligned with the $x$ axis (see Fig. \ref{fig:spheres_setup}). Mathematically, we can write that 
      \begin{equation}
        X_i (\nv) \equiv X(\rot{i}^{-1} \nv) \, ,
      \end{equation}
      where $\rot{i}$ represents the rotation by angles $(-\varphi_v, -\theta_i, -\varphi_i)$ successively, and $\rot{i}^{-1}$ is the corresponding inverse rotation, $(\varphi_i, \theta_i, \varphi_v)$.
      \item \textbf{Dipole weighting and radial profile step}. To extract the dipole signal aligned with the local transverse velocity,  we weight the rotated $X$ field by the cosine of the azimuthal angle, and then perform a binning operation $W$ along the $\theta$ direction to obtain a radial profile:
      \begin{equation}\label{eq:temp_dipole_radial}
        X_i(\theta_d) = \int \text d \nv' \, W(\theta' | \theta_d) \cos \varphi' \, X_i (\nv') \, ,
      \end{equation}
      where $W(\theta'|\theta_d)$ is a window function defining the binning in $\theta$, with $\theta_d$ indicating the chosen angular scale. Here we will assume simple top-hat bins, with $\theta_d$ the central angle, but other binning operators are used in other contexts (e.g. the compensated aperture photometry filter commonly used in kSZ stacking \cite{1510.06442}).
      \item \textbf{Stacking step}. The final stacking estimator is a velocity-weighted average of $X_i$ over all galaxies in the sample: 
      \begin{equation}
        \label{eq:stack}
         \bar X (\theta_d) = \frac{\sum_i w_iv_i X_i(\theta_d)}{\sum_i w_i},
      \end{equation}
      where $w_i$ are statistical weights associated with each galaxy.
    \end{enumerate}

    Searching for a dipolar modulation makes sense in the specific case of a spin-$1$ directional field, such as the transverse velocity, but one may also search for higher multipoles. For example, in the case of ``arrow-less'' spin-$2$ quantities, such as the local transverse tidal field \cite{alonsoRecoveringTidalField2016}, one may search for a quadrupolar pattern. We can thus generalise the dipole estimator to a general $s$-polar complex pattern as:
    \begin{equation}\label{eq:temp_spole_radial}
      X_i(\theta_d) = \int \text d \nv' \, W(\theta' | \theta_d) e^{is\varphi'} \, X_i (\nv').
    \end{equation}
    Dipolar stacking then corresponds to the real part of the above in the case $s=1$, while the standard (non-oriented) stacking estimator corresponds to the case $s=0$.

  \subsection{The power spectrum equivalence}\label{ssec:math.cls}
    We can manipulate Eq. \ref{eq:stack} to recover the harmonic-space equivalent of stacking. We start by performing a neutral operation, introducing an integral over the sphere where the integrand contains a Dirac delta function at the position of each source:
    \begin{equation}
      \bar X (\theta_d) = \frac{1}{\sum_iw_i} \sum_i \int \text d \nv \, \delta^{\mathcal D}(\nv - \nv_i ) \, v(\nv)\,X_{\nv} (\theta_d),
    \end{equation}
    where $X_{\nv}(\theta_d)$ is defined the same way as $X_i(\theta_d)$ in Eq. \ref{eq:temp_spole_radial} at an arbitrary point $\nv$. This allows us to define the modulus of the transverse momentum field as
    \begin{equation}
      \pi(\nv) \equiv \frac{1}{\bar{n}}\sum_i w_i\delta^{\mathcal D} (\nv - \nv_i) v(\nv),
    \end{equation}
    where $\bar{n}$ is the number densty of sources. This simplifies the stacking estimator to
    \begin{equation}
     \bar X(\theta_d) = \frac{\bar n}{\sum_iw_i} \int \text d \nv \, \pi(\nv) X_{\nv} (\theta_d) \,. 
    \end{equation}

    Substituting Eq. \ref{eq:temp_spole_radial} into the above, and expanding $X(\nv)$ in terms of its spherical harmonic coefficients $X_{\ell m}$:
    \begin{align} \nonumber
      \bar X(\theta_d) = \frac{\bar n}{\sum_iw_i} \int \text d & \nv \, \pi (\nv) \int \text d \nv' \, W(\theta'|\theta_d) e^{is\varphi'} \\ & \times \, \left[ \sum_{\ell m} X_{\ell m} Y_{\ell m}(\rot{\nv}^{-1} \nv') \right]. \label{eq:cosphi_rottemp}
    \end{align}
    Note that we added a $\nv$ subscript to the rotation to highlight the fact that this rotation depends on the sky position, as well as the orientation of ${\bf v}_\perp$ at that point. Next we use the properties of the spherical harmonics under rotations to write the rotated $Y_{\ell m}$ in terms of the unrotated function and the Wigner $D$ matrices of the rotation angles (see Eq. \ref{eq:ylm_rotated} in Appendix \ref{app:WignerDmatrices}). The estimator becomes
    \begin{align} 
      \nonumber \bar X(\theta_d) & = \frac{\bar n}{\sum_iw_i} \sum_{\ell m} X_{\ell m} \int \text d \nv \, \pi(\nv) \sum_{m'}D_{mm'}^{\ell*}(\varphi,\theta,\varphi_v) \\
      & \hspace{12pt}\times \int \text d \nv' \, W(\theta'|\theta_d) \, e^{is\varphi'} Y_{\ell m'}(\nv')
    \end{align}

    Now let us focus on the integral over $\nv'$. The spherical harmonics are separable according to
    \begin{equation}
      Y_{\ell m} (\theta, \varphi) = c_{\ell m} P_\ell^m(\cos \theta) e^{im\varphi},
    \end{equation}
    with $P_\ell^m$ the associated Legendre polynomials, and $c_{\ell m}$ a normalisation constant given by
    \begin{equation}\label{eq:clm}
      c_{\ell m} = (-1)^m\sqrt{\frac{2\ell +1}{4\pi} \frac{(\ell - m)!}{(\ell + m)!}}.
    \end{equation}
    The integral over $\nv'$ is then given by:
    \begin{equation}
      \int {\rm d}\nv'\,W(\theta'|\theta_d)e^{is\varphi'}Y_{\ell m'}(\nv)=\tilde{f}_{\ell,-s}(\theta_d)\,\delta^K_{m',-s},
    \end{equation}
    where $\delta^K_{m,m'}$ is the Kronecker, enforcing $m'=-s$, and we have defined
    \begin{equation}
      \tilde{f}_{\ell,m}(\theta_d)\equiv 2\pi\int {\rm d}\theta\,\sin\theta\,c_{\ell m}P_\ell^m(\cos\theta)\,W(\theta|\theta_d).
    \end{equation}

    Substituting this result, and using the separability of the $D$ matrix in its dependence on $\varphi_v$ (see Eq. \ref{eq:wignerd_exp}), we can write:
    \begin{align}\nonumber
      \bar{X}(\theta_d)=&\frac{\bar{n}}{\sum_iw_i}\sum_{\ell m}X_{\ell m}\tilde{f}_{\ell,-s}(\theta_d)\\&\times\int {\rm d}\nv\,\pi(\nv)e^{-is\varphi_v}D^{\ell*}_{m-s}(\varphi,\theta,0).
    \end{align}
    The $D$ matrices may be expressed in terms of spin-weighted spherical harmonics (Eq. \ref{eq:wignerd_swsh}), which allows us to write:
    \begin{equation}\nonumber
      \bar{X}(\theta_d)=-\frac{\bar{n}}{\sum_iw_i}\sum_{\ell m}X_{\ell m}\,_s\pi_{\ell m}^*(-1)^s\sqrt{\frac{4\pi}{2\ell+1}}\tilde{f}_{\ell,-s}(\theta_d),
    \end{equation}
    where we have defined the harmonic coefficients of the complex spin-$s$ momentum field $_s\pi(\nv)\equiv\pi(\nv)e^{is\varphi_v}$:
    \begin{align}\nonumber
      _s\pi_{\ell m}
      &\equiv -\int d\nv\,\pi(\nv)\,e^{is\varphi_v}\,_sY^*_{\ell m}(\nv)\\
      &=\frac{1}{\bar{n}}\sum_i w_i\,\tilde{v}_i\,_sY^*_{\ell m}(\nv_i)
    \end{align}
    We can simplify things further by using Eq. \ref{eq:clm} and Eq. \ref{eq:smalld_alp}, to obtain:
    \begin{equation}
      \bar{X}(\theta_d)=-\frac{\bar{n}}{\sum_iw_i}\sum_{\ell m}X_{\ell m}\,_s\pi^*_{\ell m}\,W^s_\ell(\theta_d),
    \end{equation}
    where we have defined the \emph{spin-$s$ window function} $W^s_\ell(\theta_d)$:
    \begin{equation}
      W^s_\ell(\theta_d)\equiv2\pi\int d\theta\,\sin\theta\,W(\theta|\theta_d)\,d^\ell_{s0}(\theta),
    \end{equation}
    with $d^\ell_{mm'}(\theta)$ the Wigner small-$d$ matrices (see Eq. \ref{eq:smalld_alp}). Finally, for a survey covering a sky fraction $f_{\rm sky}$, the number density is $\bar{n}=\sum_i w_i/(4\pi f_{\rm sky})$, allowing us to obtain our final expression:
    \begin{equation}\label{eq:stack_as_fct_of_cls}
      \bar{X}(\theta_d)=-\sum_{\ell}\frac{2\ell+1}{4\pi}W^s_\ell(\theta_d)\,{\cal C}^{\pi X}_\ell,
    \end{equation}
    where ${\cal C}^{\pi X}_\ell$ is the pseudo-$C_\ell$ of the $X$ and $_s\pi$ fields corrected for $f_{\rm sky}$:
    \begin{equation}
      {\cal C}^{\pi X}_\ell\equiv\frac{1}{(2\ell+1)f_{\rm sky}}\sum_m X_{\ell m}\,_s\pi^*_{\ell m}={\cal C}^{X\pi_E}_\ell-i{\cal C}^{X\pi_B}_\ell.
    \end{equation}
    Here, we have decomposed $_s\pi$ into its $E$- and $B$-mode components (see e.g. \cite{1967JMP.....8.2155G,1809.09603,2410.07077})
    \begin{equation}
      _s\pi_{\ell m}=\pi^E_{\ell m}+i\pi^B_{\ell m}.
    \end{equation}

    Having introduced the complex $e^{is\varphi}$ factor instead of the real $\cos s\varphi$ needed for extracting $s$-pole signals, the final estimator in Eq. \ref{eq:stack_as_fct_of_cls} is more general than the estimators presented in \cite{caiDetectionCosmologicalDipoles2025, lokkenSuperclusteringAtacamaCosmology2022, hadzhiyskaTracingCosmicGas2024}. We can see that the latter can be recovered from the real part of our estimator, and is solely sourced by the $E$-mode of the spin-$s$ momentum field, whereas the imaginary part (corresponding to sine-weighted stacking) is sourced by the parity-odd $B$-modes. This generalises the result presented in \cite{harscouetKSZEveryonePseudoCl2025} for non-oriented stacks.

    We thus see that multi-polar directional stacks may be recovered, without loss of information, from the cross-spectrum between the target field $X$ and a momentum field $_s\pi$ constructed by weighting each source by the value of the spin-$s$ field $\tilde{v}$ defining the preferred direction. This field, and its power spectra, may be constructed using catalogue-level techniques that exploit spherical harmonic transforms on irregular grids \cite{2304.10431,2312.12285,2407.21013}, avoiding finite-resolution effects.

    In some stacking analyses it is common to simply orient the external map along the preferred direction, without weighting the result by the value of the field defining this direction. In our formalism this can be achieved by simply replacing the spin-$s$ field $\tilde{v}=v\,e^{is\varphi_v}$ by its phase $e^{is\varphi_v}$. This is obvious in the case $s=1$ (e.g. for velocity-oriented dipoles), but the result is fully general, and Appendix \ref{app:hessianfil} shows this for $s=2$.

\section{Validation and applications}\label{sec:ex}
  In this section we validate the connection between directional stacking and power spectra numerically, and showcase the application of our harmonic-space estimator to real datasets previously studied in the literature in the context of directional stacking.

  For a given power spectrum measurement, we compute its signal-to-noise ratio (SNR) as \citep{harscouetConstraintsCMBLensing2026}
  \begin{equation}\label{eq:snr}
    \mathrm{SNR} = \sqrt{\mathbf d \mathsf C^{-1} \mathbf d^T - N_\mathrm{dof}} \,, 
  \end{equation}
  where $\mathbf d$ represents the data vector (i.e. the measured power spectrum), and $N_\mathrm{dof}$ is the number of degrees of freedom (here, the length of the data vector). $\mathsf C$ is the covariance matrix of the data vector, estimated using the usual iNKA approach of \citep{2010.09717} for power spectrum covariances. In the case of real-space stacking, we use bootstrap covariances computed from 1000 samples with replacement of the individual profiles.

  \subsection{Validation}\label{ssec:ex.val}
  
\begin{figure*}
    \centering
    \includegraphics[width=\linewidth]{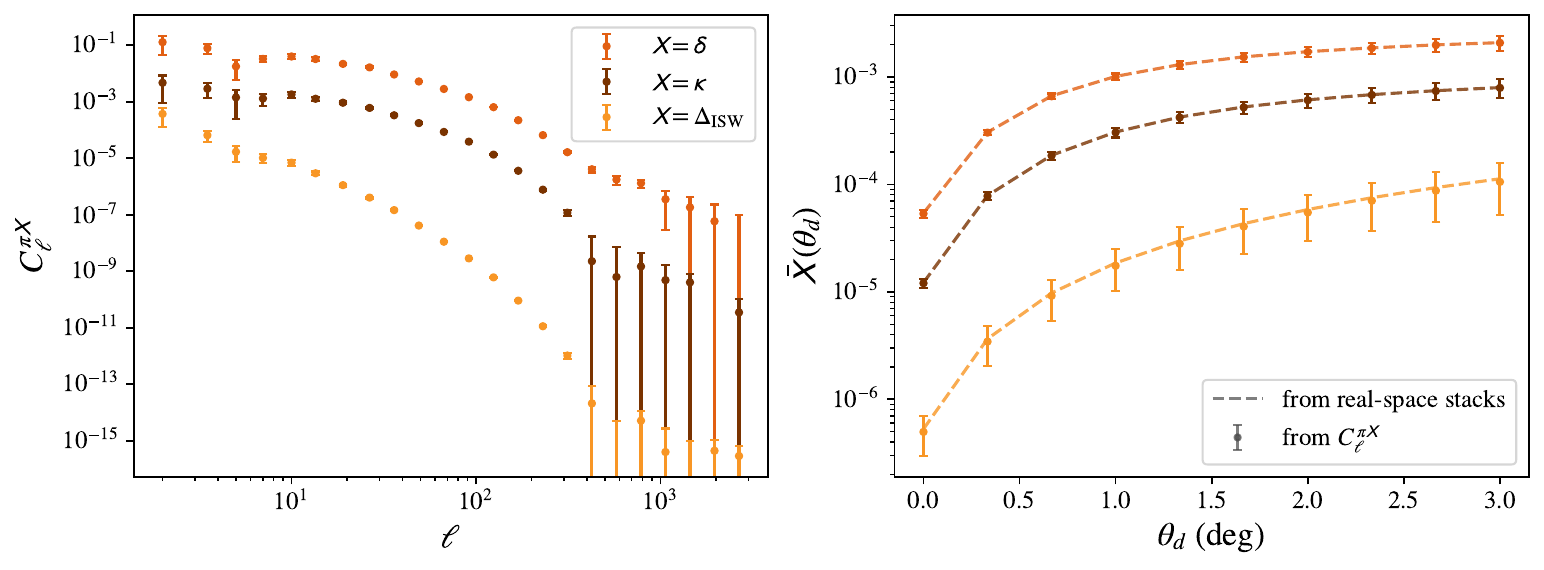}
    \includegraphics[width=\linewidth]{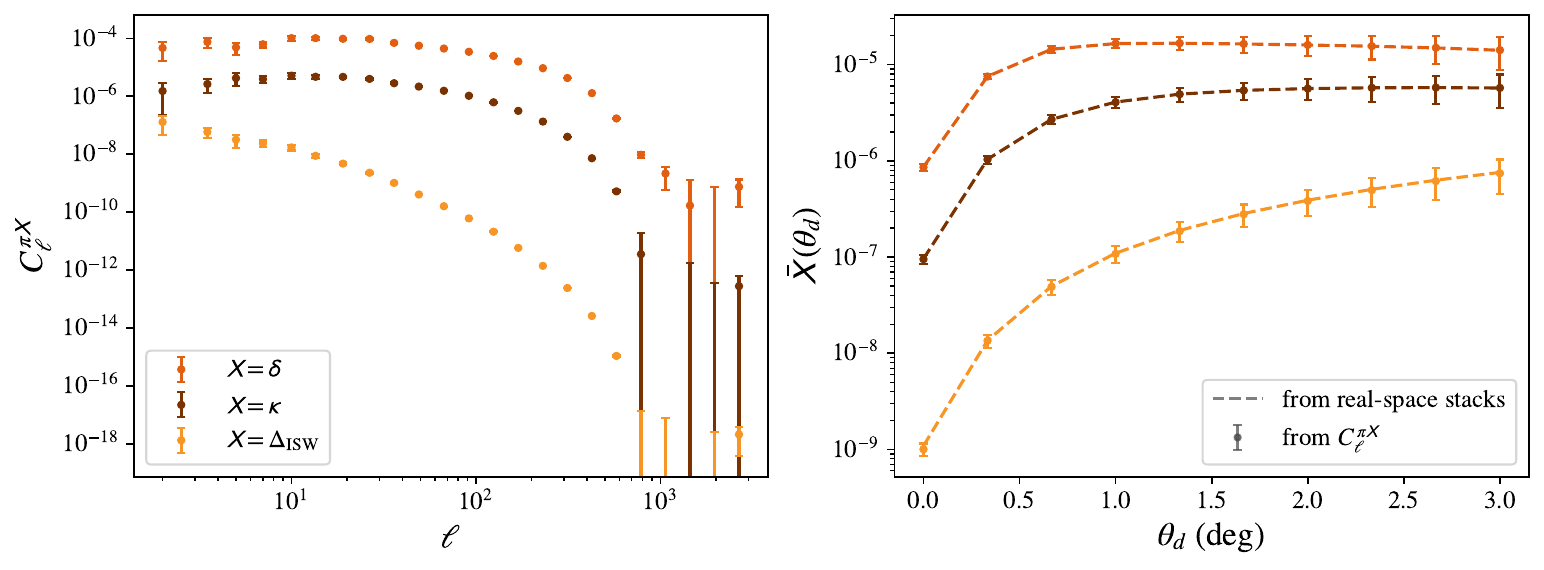}
    \caption{{\bf Top panels:} dipole ($s=1$) measurement of the correlation between the simulated galaxy overdensity ($\delta$), lensing convergence ($\kappa$) and ISW maps and the transverse galaxy momentum $_1\pi$ constructed using reconstucted velocities. The \textit{left panel} shows the result of our harmonic-space estimator. All cross-spectra with the $E$-mode of $_1\pi$ are detected at high significance, while the $B$-mode spectra are negligibly small. The \textit{right panel} shows the real-space dipole stacks using the reconstructed transverse velocity, together with their values reconstructed from the harmonic-space measurements using Eq. \ref{eq:stack_as_fct_of_cls}. {\bf Bottom panels:} quadrupole ($s=2$) measurements using the transverse tidal field reconstructed from the galaxy overdensity to define the local preferred direction. As in the $s=1$ case, all $E$-mode spectra are detected at high significance, and can be used to reconstruct the real-space stacks accurately. Note that all measurements shown in the right column were rescaled by arbitrary factors in order to display them on the same dynamic range, and their errors enhanced by arbitrary factors (top: $\times 5$, bottom: $\times 15$) to improve readability. }
    \label{fig:sim_results}
\end{figure*}

    We begin by validating the results presented in the previous section using synthetic data. We use {\tt CoLoRe}\footnote{\url{https://github.com/damonge/CoLoRe}} \citep{ramirez-perezCoLoReFastCosmological2022} to generate a mock galaxy sample covering the redshift range $z\lesssim0.8$ with a redshift-dependent number density
    \begin{equation}
      n(z)\propto \frac{1}{2}\left[1-{\rm erf}\left(\frac{\chi(z)-\chi(z_*)}{\Delta\chi}\right)\right],
    \end{equation}
    where $\chi(z)$ is the comoving radial distance to redshift $z$, $z_*=0.7$, and $\Delta\chi=430\,{\rm Mpc}$. The catalogue contains approximately $10^7$ objects distributed across the full celestial sphere, and we assume a linear bias $b=1$ for simplicity. Additionally we generate maps of the lensing convergence and the ISW effect for a source plane at redshift $z_s=1$.

    We reconstruct the transverse peculiar velocity field from the simulated galaxy catalog using {\tt pyrecon}\footnote{\url{https://github.com/cosmodesi/pyrecon}} \cite{paillasOptimalReconstructionBaryon2025, chenExtensiveAnalysisReconstruction2024}. The local transverse tidal field is reconstructed using the method of \cite{alonsoRecoveringTidalField2016}, described in Appendix \ref{app:hessianfil}, using a Gaussian smoothing of the galaxy density field with a smoothing scale of 0.3$^\circ$. We use these fields to define the preferred stacking direction, and to construct the associated momentum fields in the harmonic-space picture. They allow us to validate our results in the case of dipoles (transverse velocities, $s=1$), and quadrupoles (transverse tidal field, $s=2$).

\newpage
    To calculate the harmonic-space cross-spectra $C_\ell^{T\pi_{E/B}}$, we use {\tt NaMaster}\footnote{\url{https://github.com/LSSTDESC/NaMaster}}, taking advantage of the catalog-based formalism described in \cite{2407.21013}. Specifically, we construct the momentum fields as {\tt NmtCatalogMomentum} objects, using the discrete source positions and the values of the directional fields sampled at them. We then compute the cross-spectrum between this field and the simulated lensing convergence and ISW maps ($\kappa$ and $\Delta_{\rm ISW}$, respectively), as well as a map of the projected galaxy overdensity itself, $\delta_g$. The $E$-mode power spectrum measurements are shown in the left panels of Fig. \ref{fig:sim_results} for spin-1 and spin-2 stacking (top and bottom panels, respectively). The cross-correlations with the $E$-mode component of the transverse momentum field are detected at high significance in all cases, whereas the $B$-mode component is strongly suppressed. This is as expected, given that the reconstructed velocities and tidal field are both related to transverse derivatives of scalar quantities, and hence are dominated by $E$-modes.

    We compute the real-space dipole and quadrupole directional stacks of the same maps ($\kappa$, $\Delta_{\rm ISW}$ and $\delta_g$) using the reconstructed velocities and tidal field, applying the real-space estimator in Eq. \ref{eq:temp_spole_radial} for $s=1$ and 2 respectively. The right panels of Fig. \ref{fig:sim_results} show the resulting measurements, together with their values constructed from the power spectrum measurements, shown in the left panels, using Eq. \ref{eq:stack_as_fct_of_cls}. The real-space stacks are accurately reproduced from the power spectra, demonstrating the equivalence of both approaches at the estimator level derived in the previous section.

\begin{figure}
    \centering
    \includegraphics[width=\linewidth]{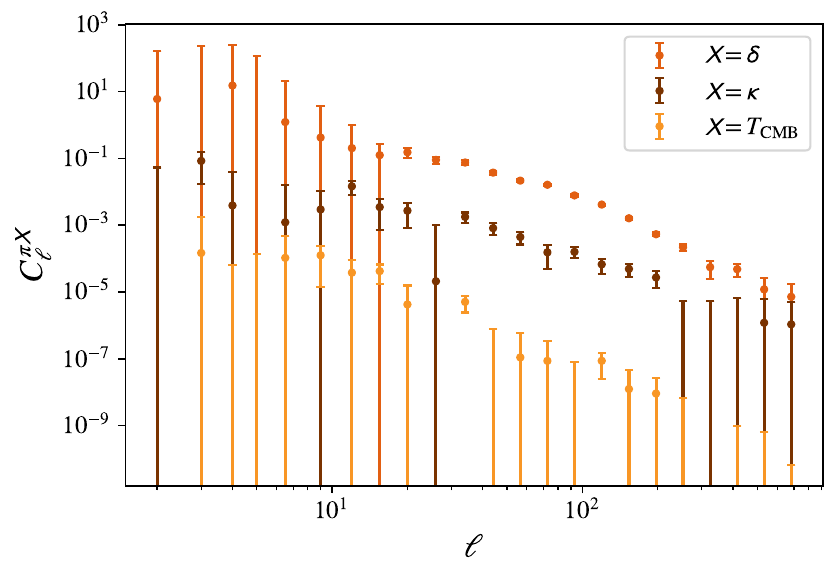}
    \caption{\textbf{Dipole ($s=1$) measurement.} The $\pi \times \delta$, $\pi \times \kappa$, and $\pi \times T_\text{CMB}$ $E$-mode power spectra measured on BOSS and Planck data, with the transverse momentum $\pi$ constructed from the reconstructed galaxy velocity field. The SNRs are $\sim 35$ and $\sim 6$ for $\delta$ and $\kappa$ respectively, while the cross-correlation with $T_{\rm CMB}$ remains undetected. The $B$-mode power spectra are consistent with 0.}
    \label{fig:data_vdipoles}
\end{figure}

\subsection{Example 1: velocity-oriented dipoles}\label{ssec:ex.spin1}
  In this first example, we mirror the analysis of \cite{caiDetectionCosmologicalDipoles2025}, where the dipolar signature of various LSS tracers around galaxies was measured, using the reconstructed transverse galaxy velocities as a directing field. For this, we use the CMASS galaxy sample of the Baryon Oscillation Spectroscopic Survey (BOSS) 12th data relsease (DR12) \cite{bosscmass}, as well as maps of the CMB primary temperature anisotropies and the lensing convergence from the \planck PR3 release \cite{planckoverview,planckgl}. We only consider galaxies in the North BOSS patch. The galaxy velocities are recovered by performing velocity reconstruction using the multigrid algorithm implemented in the \texttt{pyrecon} package. The resulting velocity catalogue is different from the one used in the original work of \cite{caiDetectionCosmologicalDipoles2025}, but we verified that we obtain compatible measurements. As shown in \cite{caiDetectionCosmologicalDipoles2025}, the most likely cosmological contribution to a dipole signature in the CMB temperature map is the ISW effect, which traces the gravitational potential field driving the peculiar motion of the CMASS galaxies. We note, however, that we expect any ISW signature to be significantly fainter than that studied in the simulations of Section \ref{ssec:ex.val}, due to the contribution from uncorrelated primary anisotropies and instrumental noise. In addition to the $s=1$ signature in the CMB temperature and convergence maps, we also study the cross-correlation with the galaxy overdensity map itself.

  From Section \ref{ssec:ex.val} we expect the signal to be most significant on large scales, so we construct low resolution maps ($N_\text{side} = 256$) of the $\delta_g$, $\kappa$ and $T_\text{CMB}$ fields, and measure the power spectra in 23 logarithmically-spaced bins between $\ell=2$ and $\ell_\text{max} = 3N_\text{side}$. As before, the transverse momentum field was constructed from the galaxy positions and reconstructed velocities, using the {\tt NmtFieldCatalogMomentum} object in {\tt NaMaster}. Power spectrum uncertainties were calculated using the analytical Gaussian approximation of \cite{2010.09717}. The resulting power spectrum measurements are shown in Fig. \ref{fig:data_vdipoles}. The $s=1$ signature in the galaxy overdensity and lensing convergence fields are detected at high significance (SNR $\sim 35$ and 6 respectively), and we do not detect a significant cross-correlation against $T_{\rm CMB}$ (${\rm SNR}\sim1$), although this was detected in the real-space analysis of \cite{caiDetectionCosmologicalDipoles2025}. The absence of signal might be due to differences in the reconstructed velocities used here, or in the estimated measurement uncertainties (we use an analytical Gaussian covariance, compared with the simulation-based estimate in \cite{caiDetectionCosmologicalDipoles2025}). It is worth noting that the real-space stacking measurements we recover with our data (and which we can also reproduce from our power spectrum measurements) seem qualitatively similar to the measurements presented in \cite{caiDetectionCosmologicalDipoles2025}.

  \newpage

  \begin{figure}
    \centering
    \includegraphics[width=\linewidth]{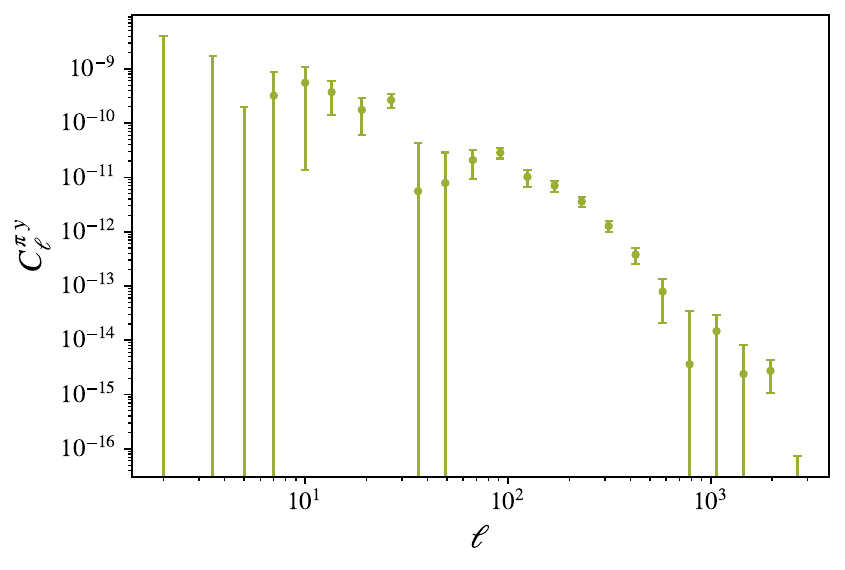}
    \caption{\textbf{Quadrupole ($s=2$) measurement.} The $\pi \times \text{tSZ}$ $E$-mode power spectrum measured on BOSS and Planck data. Here $\pi$ was constructed from the reconstructed tidal tensor with a smoothing scale of $\sim 0.3 ^\circ$. The signal is detected at SNR $\sim 10$. The B-mode spectrum is consistent with 0.}\label{fig:data_fquadpole}
  \end{figure}
  
  \subsection{Example 2: filament-oriented quadrupoles}\label{ssec:ex.spin2}
    In this second example, we aim to measure the quadrupolar signature imprinted by the filamentary structures of the cosmic web on the thermal Sunyaev-Zel'dovich (tSZ) effect, in a similar spirit to the detection of the directional tSZ signal reported in \cite{lokkenSuperclusteringAtacamaCosmology2022}. 

    The formalism needed to measure this signal follows the same structure described in Section \ref{sec:math} -- with a few additional considerations: first, the preferred direction is now informed by the direction of the local tidal field (an "arrow-less" spin-2 vector field), and second, the stacked measurement presented in \cite{lokkenSuperclusteringAtacamaCosmology2022} are not weighted by the value of the tidal field, but only oriented along its direction. As discussed at the end of Section \ref{ssec:math.cls}, this means that the directing field is not the projected tidal tensor itself, but only its phase (i.e. with unit modulus) \footnote{As a side note, another filament-aligned detection was presented in \cite{hadzhiyskaTracingCosmicGas2024} for the kSZ effect. The spin-2 field of interest in this case is a composite field, with modulus given by the amplitude of the radial velocity field, and direction given by the tidal field.}. Further information regarding the definition and spin nature of the projected tidal field is provided in Appendix \ref{app:hessianfil}.

    We use the same galaxy sample (CMASS North) employed in Section \ref{ssec:ex.spin2}, and the 2015 \planck Compton-$y$ (tSZ) map from \cite{plancktsz}. Specifically, we use the map constructed using the MILCA component separation method. Note that both datasets are different from the data used in \cite{lokkenSuperclusteringAtacamaCosmology2022}, and therefore we are not attempting to reproduce their results. The transverse tidal field was reconstructed from the galaxy overdensity using the method of \cite{alonsoRecoveringTidalField2016}, described in Appendix \ref{app:hessianfil}. The overdensity field was initially smoothed using a $\sigma = 0.3^\circ$ Gaussian kernel (corresponding to a comoving scale of about $12 \text{ Mpc}$ at the mean redshift of the CMASS sample). For this measurement we use higher-resolution maps with $N_\text{side} = 1024$, as much of the tSZ signal is present on relatively small scales. We measure the $C_\ell^{\pi y}$ power spectrum in 23 logarithmically-spaced bins between $\ell=2$ and $\ell_\text{max} = 3N_\text{side}$, which we show in Fig. \ref{fig:data_fquadpole}. The power spectrum is detected with a SNR $\sim 10$, higher than previously reported in \cite{lokkenSuperclusteringAtacamaCosmology2022}, and likely driven by the significantly larger area used here. We only report this measurement here and leave its interpretation for future work. This would require a more thorough study of the dependence on analysis choices (e.g. tidal reconstruction method), foreground contamination, etc.

  \subsection{Information content in galaxy-based directional stacking}\label{ssec:ex.info}
    In arguably the most common form of directional stacking, using the reconstructed transverse velocities at the positions of galaxies, the spin-$s$ field used is derived from the overdensity of the same galaxies on which the external field is stacked\footnote{Schematically, the reconstructed galaxy velocity is ${\bf v}({\bf k})\propto i{\bf k}\,\delta({\bf k})/k^2$.}. This thus bears the question: is significant information contained in the resulting stacking measurements beyond that which could be obtained from the direct correlation of the external field with the galaxies themselves?

    Schematically, the transverse momentum field is
    \begin{equation}
      \pi_\perp(\nv) = \int dz\,p(z)\,\tilde{v}_{g,\perp}(\chi(z)\nv)\,[1+\delta_g(\chi(z)\nv)],
    \end{equation}
    where $p(z)$ is the redshift distribution of the galaxies under study, $\delta_g$ is the galaxy overdensity, and $\tilde{v}_{g,\perp}$ is the reconstructed velocity field (where we have added the subscript $g$ to remind ourselves that this is estimated from $\delta_g$). We can therefore split $\pi_\perp$ into two components, $\pi_\perp=\pi^{v}_\perp+\pi^{v\delta}_\perp$, where
    \begin{align}
      &\pi^v_\perp(\nv)\equiv\int dz\,p(z)\,\tilde{v}_{g,\perp}(\chi(z)\nv),\\
      &\pi^{\delta v}_\perp(\nv)\equiv\int dz\,p(z)\,(\delta_g\,\tilde{v}_{g,\perp})(\chi(z)\nv).
    \end{align}
    The cross-spectrum between $\pi_\perp$ and the external field $X$ then receives two additive contributions: $C_\ell^{v,X}\sim\langle \pi^v_\perp X\rangle$, and $C_\ell^{\delta v,X}\sim\langle \pi^{\delta v}_\perp X\rangle$. The first contribution is sensitive to the $\delta_g\times X$ power spectrum, while the second one is sensitive to the $\delta_g\times\delta_g\times X$ bispectrum. In principle, on perturbative scales the second contribution should be subdominant over the first one, since it involves a higher-order correlator. In the case of the radial momentum field (i.e. that constructed using the radial velocity), used in measurements of the kSZ effect \cite{harscouetKSZEveryonePseudoCl2025}, the first contribution is heavily suppressed, as it is a purely longitudinal mode that cancels out when integrated along the line of sight \cite{1301.3607,1506.05177,2509.18732}. Thus, kSZ measurements are in fact primarily sensitive to the bispectrum contribution. However, as we show in Appendix \ref{app:transverse_momentum}, this is not the case for transverse velocities, and both terms survive.

    The question then is whether the bispectrum contribution is sufficiently large to change the information content of $C_\ell^{\pi X}$ significantly from that contained in $C_\ell^{v,X}$, which could be estimated from the power spectrum between $X$ and a suitably weighted version of the galaxy overdensity (since $v\sim \delta_g/k$) without resorting to any directional stacking.

    \begin{figure}
      \centering
      \includegraphics[width=\linewidth]{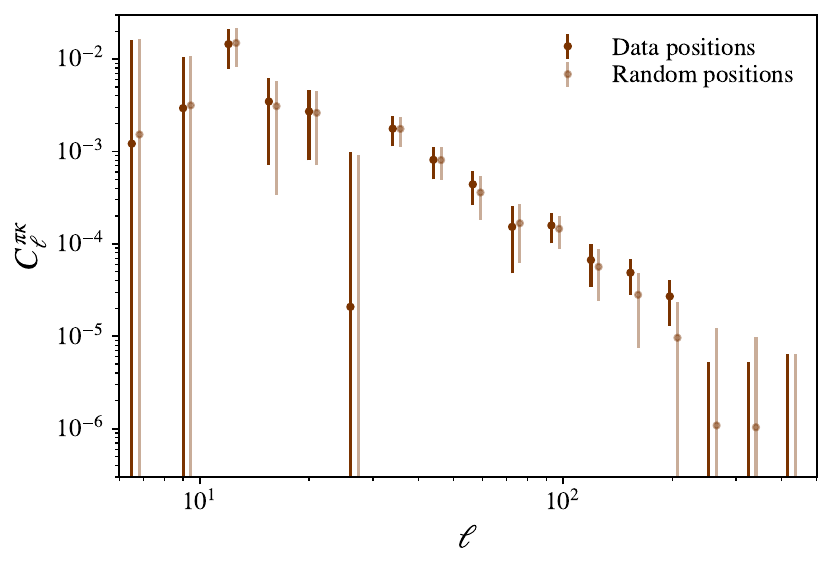}
      \caption{Cross-spectrum between the \planck CMB lensing convergence map $\kappa$ and the spin-1 transverse momentum field constructed from the velocities reconstructed using the BOSS CMASS sample, from which the lensing velocity dipole can be derived. The dark brown points show the measurements made when evaluating these velocities at the positions of the CMASS galaxies, i.e. the standard momentum field (this measurement is also shown in Fig. \ref{fig:data_vdipoles}), while the lighter points show the result of evaluating these velocities at the positions of a random catalogue. The latter measurement is only sensitive to the cross-correlation between $\kappa$ and the galaxy overdensity (the latter of the two via the reconstructed velocity field). We find that the galaxy-$\kappa$ power spectrum dominates the signal and information content, with the bispectrum-like contribution described in the main text remaining sub-dominant.}
      \label{fig:data_vs_random}
    \end{figure}
    We can address this question quantitatively using the data used to measure the lensing convergence dipole in Section \ref{ssec:ex.spin1}. Here, we can isolate the contribution from $C_\ell^{v,X}$ by constructing a transverse momentum field from the velocities reconstructed from the galaxy data but \emph{evaluated at the positions of a random catalogue}, and correlating it with the $\kappa$ map. We can then compare this with the cross-spectrum between $\kappa$ and the standard transverse momentum field (i.e. with the velocities evaluated at the real galaxy positions). Any differences between both would then be due to the bispectrum contribution $C_\ell^{\delta v,X}$. We show the result of this exercise in Figure \ref{fig:data_vs_random}, with the cross-spectra estimated using random positions and real galaxy positions shown in red and blue, respectively. We do not observe any significant difference between both measurements, demonstrating that the signal is indeed dominated by $C_\ell^{v,X}$, and thus sensitive mostly to the galaxy-$\kappa$ power spectrum. A similar result applies to the case of  quadrupole stacking based on the local tidal field when the latter is reconstructed from the galaxy overdensity.

    Analyses using more sensitive datasets may be able to detect the bispectrum term, which does contain additional information. To isolate this contribution, one may proceed as above, subtracting the signal measured using random positions, and this is indeed the procedure followed in \cite{hotinliMLinprep} to detect the moving lens effect. Finally, directional stacks using spin-$s$ fields that are not derived from the galaxy distribution will contain information complementary to the cross-correlation with galaxies, and the harmonic-space approach advocated here may be used to carry out such analyses cleanly and efficiently.

\section{Conclusions}\label{sec:conc}
  Directional stacking has become a useful tool to study interesting astrophysical signatures. By orienting the map of interest around preferred directions, aligned with key properties of the local environment, we gain access to morphological aspects of the sources under study (e.g. the distribution of gas density or other thermodynamic properties along filaments, or along preferred galaxy axes), as well as dynamical properties of the large-scale structure (LSS) (e.g. the cosmic velocity field, or gravitational lensing caused by extended structures) relevant for cosmology \cite{gallegoStackingCosmicWeb2018, lokkenSuperclusteringAtacamaCosmology2022, hadzhiyskaTracingCosmicGas2024,  caiDetectionCosmologicalDipoles2025, beheshtiMovingLensEffect2025, hotinliMLinprep, nikjooWeakLensingPhotometric2026}. Over the last few years, it has become clear in LSS analyses that combining data from multiple probes is key to breaking key parameter degeneracies and to pin down astrophysical uncertainties (e.g. to constrain clustering properties of galaxies, their intrinsic alignments, and the impact of baryonic feedback in the matter power spectrum). Since directional probes carry complementary information on both cosmology and astrophysics, folding them into such combined analyses will become increasingly important, as their sensitivity improves with the higher-quality data of ongoing and future experiments.

  In this paper we have contributed towards this effort, putting forward a harmonic-space equivalent of directional stacking that offers a number of advantages over the real-space approach. We have shown that all the information present in directional stacks may be recovered from the cross-correlation between the target map and the ``transverse momentum'' field constructed by weighting each source by the value of the spin quantity used to define the preferred local direction. Typical choices for this directing field are the spin-1 reconstructed transverse velocities, or the spin-2 tidal tensor. The $E$- and $B$-mode contributions to this cross-correlation can then be related to the parity-even and odd components of the stacked signal without loss of information. This approach carries the same advantages of other power spectrum approaches: measurements on different harmonic scales are largely uncorrelated, allowing for an easier separation between different angular scales and for fast assessments of detection significance and goodness of fit. Power spectrum estimators are typically much faster than real-space correlations, and their statistical uncertainties may be estimated quickly and accurately using existing analytical methods \cite{1906.11765,2010.09717}. The resulting power spectrum measurements can thus be easily folded into a joint-probes analysis seamlessly and self-consistently. 

  We have demonstrated the equivalence between the harmonic-space estimator and directional stacking in terms of information content, both mathematically and numerically. We have also shown its applicability to real data, measuring the dipolar signature of the galaxy overdensity and the CMB lensing convergence aligned with the reconstructed transverse velocities of galaxies, as well as the quadrupolar signature of the SZ effect around filaments aligned with the transverse tidal forces. We have also argued that, in directional stacking analyses where the directing field is reconstructed from the galaxy overdensity itself, the dominant contribution to the measurement is driven simply by the power spectrum between the target map and the galaxy overdensity itself, although additional information may be extracted if the additional bispectrum contribution discussed in Section \ref{ssec:ex.info} can be isolated (e.g. by subtracting the correlation with the directing field sampled at random positions).

  Further information may be gained by employing directing spin fields that are independent from the galaxy overdensity. Examples of these could be the measured galaxy shapes, the strength and direction of jets in radio galaxies, their polarised intensity, the spins of galaxies, groups, or clusters, or even measurements of the true transverse velocities in proper motion surveys \citep{2305.15893}. The estimator presented here, and implemented in the public code {\tt NaMaster}, may be used to measure these signatures, and to seamlessly combine them with standard LSS probes in joint analyses.

\section*{Acknowledgments}
  We would like to thank Shadab Alam, Yan-Chuan Cai, and John Peacock for useful discussions, and for making the data used in \cite{caiDetectionCosmologicalDipoles2025} available to us to validate some of our measurements. As we were finalising this manuscript, we learnt of the work of \cite{hotinliMLinprep}, where a framework similar to that described here is proposed. LH is supported by a Hintze studentship, which is funded through the Hintze Family Charitable Foundation. AW is supported by a Science and Technology Facilities Council (STFC) studentship. We made extensive use of computational resources at the University of Oxford Department of Physics, funded by the John Fell Oxford University Press Research Fund.

\onecolumngrid

\appendix

\section{Rotations and Wigner $D$-matrices}\label{app:WignerDmatrices}
  In this appendix, we collect all properties of the Wigner $D$-matrices that are relevant for the derivations of Section \ref{sec:math}. Further details may be found in e.g. \cite{varshalovichQuantumTheoryAngular1988, alma990103549800107026}.
  
  Any rotation on the sphere can be described by three successive rotations along pre-defined axes. The effect of successive rotations by angles $\gamma$, $\beta$, and $\alpha$ around the $z$, $y$ and $z$ axes respectively ($z-y-z$ convention), can be represented by the rotation operator $\hat D(\alpha, \beta, \gamma)$ -- note how the first rotation by angle $\gamma$ is written last in this notation. Also note the correspondence with spherical coordinates (see Fig. \ref{fig:spheres_setup} for easier visualisation): the azimuthal angle $\varphi$ can define a rotation about the $z$ axis, while the colatitude $\theta$ is a rotation about the projection of $\nv$ in the $x-y$ plane (if $\nv$ is aligned with the $x$ axis, it is effectively defining a rotation about the $y$ axis). The inverse rotation $\hat D^{-1}$ can be obtained by swapping the first and last rotations, and taking the negative of the original angles: 
  \begin{equation}
    \hat D^{-1}(\alpha, \beta, \gamma) \equiv \hat D(-\gamma, -\beta, -\alpha).
  \end{equation}
  From Fig. \ref{fig:spheres_setup}, it is evident that going from (4) to (1) can be achieved via the successive rotations of first $+\varphi_v$, then $+\theta$ and $+\varphi$ about $z$, $y$ and $z$ respectively -- or in our notation, $\hat D(\varphi, \theta, \varphi_v)$.
  
  The effect of this rotation operator $\hat D$ on the spherical harmonics (SH) can then be written in a matrix form, often called the Wigner D-matrix $D_{mm'}^\ell$:
  \begin{equation}\label{eq:ylm_rotated}
    Y_{\ell m} (\nv') = \sum_{m'} \left[D_{mm'}^\ell (\alpha, \beta, \gamma) \right]^* Y_{\ell m'}(\nv).
  \end{equation}
  I.e. the rotated SHs are a linear combination of original SHs of the same order $\ell$. Here, the two coordinates $\nv$ and $\nv'$ are related by a rotation: $\nv \equiv \hat D \nv'$.

  The Wigner D-matrix can be decomposed into different angular components:
  \begin{equation}\label{eq:wignerd_separable}
    D_{mm'}^\ell(\alpha, \beta, \gamma) = e^{-im\alpha} d^{\ell}_{mm'} (\beta) e^{-im' \gamma}. 
  \end{equation}
  Here we have introduced the small $d$-matrix $d^{\ell}_{mm'}$, which contains the $\beta$-dependent part of the Wigner $D$-matrix. For $m' = 0$, this small $d$ matrix has a very simple expression: 
  \begin{equation}\label{eq:smalld_alp}
    d_{m0}^{\ell }(\beta)={\sqrt {\frac {(\ell -m)!}{(\ell +m)!}}}\,P_{\ell }^{m}(\cos \beta)=\sqrt{\frac{(\ell+m)!}{(\ell-m)!}}P_\ell^{-m}(\cos\beta).
  \end{equation}
  where $P_\ell^m$ is an associated Legendre polynomial.

  From Eq. \ref{eq:wignerd_separable}, we can derive a simple property which will prove useful in Section \ref{sec:math}:
  \begin{equation}\label{eq:wignerd_exp}
    D_{mm'}^\ell (\alpha, \beta, \gamma) \equiv D_{mm'}^\ell (\alpha, \beta, 0) e^{- i m' \gamma} = D_{mm'}^\ell (0, \beta, \gamma) e^{- i m \alpha} \,.
  \end{equation}

  In the flat-sky limit, i.e. when $\ell \gg m, m'$ for a small-$d$ matrix $d_{mm'}^\ell$, the small-$d$ matrix can be related to the Bessel functions as 
  \begin{equation}
    \label{eq:smalld_to_bessel}
    d_{mm'}^\ell (\beta) \approx J_{m-m'}(\ell \beta). 
  \end{equation}

  As mentioned above, one can liken some of the angles $\alpha, \beta, \gamma$ in the $z-y-z$ convention to the spherical coordinates angles $\theta, \varphi$. Specifically, we can express the D-matrix using spin-weighted spherical harmonics (SWSH) \cite{astro-ph/0303414}: 
  \begin{equation}
    \label{eq:wignerd_swsh}
    D_{-m s}^\ell(\varphi, \theta, 0)=(-1)^m \sqrt{\frac{4 \pi}{2 l+1}}{ }_s Y_{\ell m}(\nv)~. 
  \end{equation}
  where $\nv \equiv (\sin\theta\cos\varphi,\sin\theta\sin\varphi,\cos\theta)$.

\section{Reconstructing the projected tidal field}\label{app:hessianfil}
  To recover the direction of density ridges in the cosmic web, we apply the method presented in \cite{alonsoRecoveringTidalField2016} and previously used in directional stacking works \cite{hadzhiyskaTracingCosmicGas2024, lokkenSuperclusteringAtacamaCosmology2022}. The method is based on estimating the projected tidal tensor, which determines how tidal forces deform extended objects. The eigenvalues and eigenvectors of the tensor can then be used to find the directions of maximum compression or stretching, which align with the local direction of filaments in the projected cosmic web. The 2D tidal tensor field $t_{ab}$ is defined as the covariant Hessian of the 2D potential $\phi$: 
  \begin{equation}
    t_{ab} = \mathsf H_{ab} \phi, \quad\text{where } \quad \mathsf{H} \equiv \begin{pmatrix} \partial_\theta^2 & \partial_\theta (\partial_\varphi / \sin\theta) \\ \partial_\theta (\partial_\varphi / \sin\theta) & \partial_\varphi^2 / \sin^2\theta + \cot \theta \partial_\theta \end{pmatrix}.
  \end{equation}
  The potential $\phi$ is simply extracted from the projected density field $\delta$ inverting the Poisson equation on the sphere $\nabla^2_\theta\phi=\delta$ (or, in harmonic space, $\delta_{\ell m}=-\ell(\ell+1)\phi_{\ell m}$.

  $t_{ab}$ may be written in terms of the spin-raising operator $\eth$ acting on $\phi$ (see \cite{1967JMP.....8.2155G,alonsoRecoveringTidalField2016}) as
  \begin{equation}
    {\sf t}=\frac{1}{2}\left(
    \begin{array}{cc}
      \delta + \tau_1 & \tau_2 \\
       \tau_2  & \delta-\tau_1
    \end{array}
    \right)\equiv\frac{1}{2}\mathbb{I}\delta+\frac{1}{2}\hat{\tau},\hspace{12pt}\delta\equiv\bar{\eth}\eth\phi,\hspace{12pt}\tau_1+i\tau_2\equiv\eth\eth\phi,
  \end{equation}
  where $\mathbb{I}$ is the $2\times2$ identity. The overdensity field $\delta$ is evidently a scalar under rotations, given by the trace of ${\sf t}$, whereas the traceless component of ${\sf t}$, $\hat{\tau}$, is determined by the spin-2 field $\tau\equiv\tau_1+i\tau_2$. Since $\tau$ is a second derivative of a scalar ($\phi$), it is a pure $E$-mode field ($\tau^B=0$), with
  \begin{equation}
    \tau^E_{\ell m}=-\sqrt{\frac{(\ell+2)!}{(\ell-2)!}}\phi_{\ell m}=\sqrt{\frac{(\ell+2)(\ell-1)}{\ell(\ell+1)}}\delta_{\ell m}\simeq\delta_{\ell m},
  \end{equation}
  where the last approximate equality holds in the large-$\ell$ limit.

  The tidal tensor may be used to define the preferred directions of the local cosmic web. Specifically, its principal eigenvector (i.e. that related to its largest eigenvalue) defines the direction along which tidal forces produce the largest compression (i.e. perpendicular to the direction of filaments) \cite{alonsoRecoveringTidalField2016,hadzhiyskaTracingCosmicGas2024,lokkenSuperclusteringAtacamaCosmology2022}. The eigenvectors of ${\sf t}$ are the same as those of its spin-2 traceless component $\hat{\tau}$, and their eigenvalues are related by $\lambda_t=\lambda_\tau+\delta/2$. We can therefore restrict ourselves to using only the traceless $\hat{\tau}$ to define the local tidal direction. The eigenvalues of $\hat{\tau}$ are $\lambda_\pm=\pm\sqrt{\tau_1^2+\tau_2^2}$. Let ${\bf e}_+\equiv(\cos\varphi_t,\sin\varphi_t)$ be the normalised eigenvector corresponding to $\lambda_+$, defined by the angle $\cos\varphi_t$ with respect to the local $x$ axis (the other eigenvector is then ${\bf e}_-=(-\sin\varphi_t,\cos\varphi_t)$). The traceless tensor is then given by:
  \begin{equation}
    \hat{\tau}=\lambda_+{\bf e}_+{\bf e}_+^\dagger+\lambda_-{\bf e}_-{\bf e}_-^\dagger=\lambda_+({\bf e}_+{\bf e}_+^\dagger-{\bf e}_-{\bf e}_-^\dagger)=\lambda_+\left(
    \begin{array}{cc}
      \cos2\varphi_t & \sin2\varphi_t \\
      \sin2\varphi_t   & -\cos2\varphi_t
    \end{array}
    \right).
  \end{equation}
  Thus, the complex spin-2 field may be written as $\tau=\tau_1+i\tau_2=\sqrt{\tau_1^2+\tau_2^2}\,e^{i2\varphi_t}$. That is, in the case of tidal-field reconstruction, the angle $\varphi_v$ defined in Section \ref{ssec:math.stack} corresponds to the angle defining the local direction of the cosmic web (i.e. that defining the principal eigenvector of ${\sf t}$).


\section{The projected transverse momentum}\label{app:transverse_momentum}
  We follow a formalism similar to that of \cite{1301.3607,2509.18732} to describe the structure of the projected galaxy momentum field used in directional stacking analyses. Let us define the three-dimensional momentum field
  \begin{equation}\label{eq:3dmom}
    {\bf q}({\bf x})\equiv(1+\delta_g({\bf x}))\,{\bf v}_g({\bf x}),
  \end{equation}
  where ${\bf v}_g$ is the reconstructed galaxy velocity field. We can separate ${\bf q}$ into its ``gradient'' and ``curl'' contributions\footnote{In \cite{2509.18732} the more common nomenclature of ``longitudinal'' and ``transverse'' modes was used. We avoid this here to prevent confusion with the use of the term ``transverse'' referring to vectors perpendicular to the line of sight $\nv$.} in Fourier space, i.e. those parallel and perpendicular to the wave-vector ${\bf k}$:
  \begin{equation}
    {\bf q}({\bf k})=q_\parallel({\bf k})\,\hat{\bf k}+{\bf q}_\perp({\bf k}),
  \end{equation}
  where $q_\parallel\equiv \hat{\bf k}\cdot{\bf q}({\bf k})$, and ${\bf q}_\perp\equiv(\mathbb{I}-\hat{\bf k}\hat{\bf k}^\dagger){\bf q}({\bf k})$, with $\hat{\bf k}\equiv{\bf k}/k$. Most velocity reconstruction algorithms rely on inverting the linear continuity equation, in which case the velocity is calculated as a pure gradient of the form ${\bf v}_g({\bf k})\propto i{\bf k}\delta_g({\bf k})/k^2$. Thus, only the second-order term $\delta_g\,{\bf v}_g$ in Eq. \ref{eq:3dmom} contributes to the transverse momentum ${\bf q}_\perp$. Hence, ${\bf q}_\perp$ is, in principle, subdominant to $q_\parallel$, which is sourced by ${\bf v}_g$ at first order.

  We may now define the projected radial and transverse momentum fields, $\pi_\parallel(\nv)$ and $\boldsymbol{\pi}_\perp(\nv)$ in terms of ${\bf q}$ as:
  \begin{align}
    &\pi_\parallel(\nv)\equiv\int d\chi\,H(z)\,p(z)\,\nv^\dagger{\bf q}(\chi\nv),
    \hspace{12pt}\boldsymbol{\pi}_\perp(\nv)\equiv\int d\chi\,H(z)\,p(z)\,[\mathbb{I}-\nv\,\nv^\dagger]\,{\bf q}(\chi\nv),
  \end{align}
  where $H(z)$ is the expansion rate and $p(z)$ is the redshift distribution of the galaxies. The operator $\mathbb{I}-\nv\,\nv^\dagger$ in $\boldsymbol{\pi}_\perp$ projects ${\bf q}$ on the plane perpendicular to the line of sight, while the dot product with $\nv$ in $\pi_\parallel$ selects the radial component of ${\bf q}$. Separating ${\bf q}$ into its gradient and curl components:
  \begin{align}
    &\pi_\parallel(\nv)=\int d\chi\,H(z)\,p(z)\int \frac{d^3k}{(2\pi)^3}e^{ik\chi\,x}\left(xq_\parallel({\bf k})+\nv^\dagger{\bf q}_\perp({\bf k}),\right)\\
    &\boldsymbol{\pi}_\perp(\nv)=\int d\chi\,H(z)\,p(z)\int \frac{d^3k}{(2\pi)^3}e^{ik\chi\,x}\left(\hat{\bf k}\,q_\parallel({\bf k})-\nv\,x\,q_\parallel({\bf k})+[\mathbb{I}-\nv\nv^\dagger]{\bf q}_\perp({\bf k}),\right),
  \end{align}
  where $x\equiv\nv\cdot\hat{\bf k}$.
  
  For most scales and distances of interest, the product $k\chi\gg1$, and thus the factor $e^{ik\chi\,x}$ oscillates fast, leading to cancellations except when $x\simeq 0$ (i.e. when the Fourier mode is close to transverse to the line of sight). The terms proportional to $x$ above are therefore heavily suppressed. As shown in \cite{1506.05177,2509.18732}, this makes the contribution from the gradient mode $q_\parallel$ negligible in the case of $\pi_\parallel$, and is the reason why measurements of the kSZ effect are only sensitive to the curl component of ${\bf q}$ (i.e. the second-order term $\propto \delta_g\,{\bf v}_g$). In turn the gradient term $\hat{\bf k}\,q_\parallel({\bf k})$ survives in the transverse momentum $\boldsymbol{\pi}_\perp$ and, as we show in Section \ref{ssec:ex.info}, it dominates over the second-order curl component.

\twocolumngrid

\bibliography{main}

\end{document}